\documentstyle[12pt]{article}

\author{Hao Chen\\
Department of Mathematics\\
Zhongshan University\\
Guangzhou,Guangdong 510275\\
People's Republic of China}
\title{Quantum Entanglement  Without Eigenvalue Spectra}
\date{July,2001}

\begin{document}

\maketitle
\begin{abstract}

From the consideration of measuring bipartite mixed states by separable pure states, we introduce algebraic sets in complex projective spaces for the bipartite mixed states as the degenerating locus of the measurement. These algebraic sets  are non-local invariants independent of the eigenvalues  and only measure the position of the eigenvectors of bipartite mixed states. The algebraic sets have to be the sum  of the  linear subspaces if the mixed states are separable, and thus we give a new criterion of separability. Based on our criterion, examples are given to illustrate that entangled mixed states which are invariant under partial transposition (thus PPT bound entanglement) or fulfill entropy and disorder criterion of separability can be constructed systematically. Thus we reveal the point that a large part of quantum entanglement phenomenon is independent of eigenvalue spectra, and develop a method measuring this part of quantum entanglement. The results are also extended to multipartite case.

\end{abstract}

Quantum entanglement was first noted as a feature of quantum mechanics in the famous Einstein, Podolsky and Rosen [1] and Schrodinger [2] papers. Its importance lies not only in philosophical considerations of the nature of quantum theory, but also in applications where it has emerged recently that quantum entanglement is the key ingredient in quantum computation [3] and communication [4] and plays an important role in cryptography [5,6].\\

A mixed state $\rho$ in the bipartite quantum system $H=H_A^m \otimes H_B^n$ is called separable if it can be written in the form $\rho=\Sigma_j p_j |\psi_j><\psi_j| \otimes |\phi_j><\phi_j|$, where $p_j >0$ and $|\psi_j>,|\phi_j>$ are pure states in $H_A^m,H_B^n$. Otherwise it is called entangled. From the point view of quantum entanglement, two states are completely equivalent if one can be transformed into the other by the means of local unitary transformations (ie., $U_A \otimes U_B$ where $U_A,U_B$ are the unitary transformations of $H_A^m,H_B^n$.). It is clear that the property being separable or entangled of a mixed state  is preseved  after local unitary transformations. Thus for the puropose to quantify entanglement, any good measure of entanglement must be invariant under local transformations ([6,7,8]).\\

To find good necessary condition of separability (separability criterion) is the fundamental problem in the study of quantum entanglement ([9,10]). Bell's inequality ([9]) and entropy criterion ([10]) are well-known scalar criteria of separable states. In 1996, Peres [11] gave a striking simple criterion which asserts that a separable mixed state $\rho$ necessarily has positive partial transpose (PPT, ie., the partial transpose $\rho^{PT}$ , where $<ij|\rho|kl>=<il|\rho^{PT}|kj>$, has no negative eigenvalue), which has been proved by Horodeckis ([12]) also a sufficient condition of separability in $ 2 \times 2 $ and $2 \times 3$ systems. The significance of PPT property is also reflected in the facts that PPT mixed states satisfy Bell inequalities ([13]) and cannot be distilled(ie., no singlet can be extracted from it by local quantum operations and classical communications (LOCC), [14], bound entanglement)). In [15],[10] it was proved that the ranges of separable states $\rho$ and its partial transpose $\rho^{PT}$ are the linear span of separable pure states. Based on this so-called range criterion [10] several families of PPT entangled mixed states were constructed  in [15],[14],[16],[17],[18],[19] by proving that there is no separable pure state in the ranges of the constructed mixed states (edges states). However constructing PPT entangled mixed states (thus bound entanglement) is exceedingly difficult task ([16]), and the only known systematic way of such construction is the context of unextendible product base (UPB) in [16].  In real applications of the range criterion it seems there  are some difficulties to detect those entangled states for which there are some separable pure states within their ranges. The most recent disorder criterion of separability  in [20], which is stronger than entropy criterion, was proved by the mathematics of majorization.\\

In this paper we propose the following point of view: quantum entanglement of mixed states has a quite large part which is totally independent of the eigenvalue spectra of the mixed states and only dependent on the geometric position of eigenvectors, even for mixed states with the same global and local spectra there are {\em continuous many} of them which are NOT equivalent under local unitary transformations (thus {\em continuous many distinct } entanglement with the same global and local spectra, Theorem 4 below). This point has been partially manifested in Horodecki's range criterion [15,10], UPB context in [16,17] and Examples in [20]. \\

The physical motivation is as follows. For any given bipartite mixed state $\rho$ in $H_A^m \otimes H_B^n$ , we want to understand it by measuring it with separable pure states, ie., we consider the $<\phi_1 \otimes \phi_2 |\rho|\phi_1 \otimes \phi_2>$ for any pure states $\phi_1 \in H_A^m$ and $\phi_2 \in H_B^n$. For any fixed $\phi_1 \in P(H_A^m)$, where $P(H_A^m)$ is the projective space of all pure states in $H_A^m$,  $<\phi_1 \otimes \phi_2 |\rho|\phi_1 \otimes \phi_2>$ is a Hermitian bilinear form on $H_B^n$, denoted by $<\phi_1|\rho|\phi_1>$ . We consider the {\em degenerating locus } of this bilinear form, ie., $V_A^k(\rho)=\{\phi_1 \in P(H_A^m): rank (<\phi_1|\rho|\phi_1>) \leq k\}$ for $k=0,1,...,n-1$. It is proved that these {\em degenerating locus } are algebraic sets (ie., zero locus of several homogeneous multi-variable polynomials ) in $P(H_A^m)$, which are independent of the global spectra of the mixed states and only measure the geometric position of eigenvectors in $H_A^m \otimes H_B^n$ . These algebraic sets are invariants of the mixed states under local unitary transformations, and thus many numerical algebraic-geometric invariants (such as dimensions, number of irreducible components) and Hermitian differential geometric invariants (with the Fubini-Study metric of complex projective spaces, such as volumes, curvatures) of these algebraic sets are automatically invariant under local unitary transformations. In this way many candidates for good entanglement measure or potentially entanglement monotone independent of global spetra of mixed states are offerd. Another important aspect is that these algebraic sets can be easily calculated and thus can be easily used to treat the problem of quantum entanglement.\\

Based on these algebraic sets we prove a new separability criterion (independent of global spectra) which asserts that the algebraic sets have to be the sum of linear subspaces of $P(H_A^m)$ if the mixed state is separable.  Based on this new separabilty criterion, entangled mixed states fulfilling entropy and disorder criteria of separability can be constructed easily, and a continuous family of isospectral (the same global and local spectra) entangled mixed states whose members are not equivalent under local unitary transformations is constructed (Example 1). Thus it is clear that our criterion is not equivalent to the entropy and disorder criteria and can be used to detect entangled states not violating these two criteria. We also illustrate a systematic way for constructing  PPT (actually $\rho^{PT}=\rho$) entangled mixed states (thus bound entanglement) in Example 2. There are some separable pure states in the ranges of the entagled states constructed in Example 2, thus it seems not so direct and easy to detect these entangled states by the the range criterion. Our criterion based on algebraic-geometric invariants of mixed states works well to detect those entangled states for which there are some separable pure states within their ranges.\\

Although the conclusion that the ranges of separable mixed states are the linear span of separable pure states can imply our separability criterion Theorem 3, the criterion here can be used more effectively than the range criterion in the case that there are some separable pure states within the ranges of the entangled mixed states, because in this case it is not so direct and easy to prove the ranges of these entangled mixed states are not linear span of separable pure states and thus get a contradiction to the range criterion, however we can easily compute the algebraic-geometric invariants of these entangled mixed states and use the systematic results in  algebraic geometry to prove that these invariants are not linear even when there are some separable pure states in the ranges of these constructed entangled mixed states(see Example 2). In many cases it is almost immediately clear from our separability criterion Theorem 3 here that some mixed states are entangled since their corresponding algebraic-geometric invariants are clearly {\em curved} and thus not linear (see Example 1). On the other hand all known PPT bound entangled mixed states were constructed by proving there is no separable pure state in their ranges (see [15],[14],[16],[17],[18],[19]). Our criterion offered a new systematic way to construct MORE families of PPT bound enatngled mixed states by choosing some matrices satisfying some special properties (see Example 2 and discussion there).\\

It is clear that any separability criterion for bipartite mixed states can be applied to multipartite mixed states for their separability under various cuts. In [21], Horodeckis proved a separability criterion for multipartite mixed states by the using of linear maps. The algebraic-geometric invariants and the separability criterion in bipartite case can be extended naturally to multipartite mixed states. We introduce algebraic sets in the products of complex projective spaces for the multipartite mixed states, which are independent of the eigenvalues of the mixed states and only measure the {\em geometric position} of eigenvectors. These algebraic sets are invariants of the mixed states under local unitary transformations. Based on these algebraic sets we prove a new separability criterion (independent of eigenvalues) which asserts that the algebraic sets have to be the sum of the products of linear subspaces if the multipartite mixed state is separable. Based on this new separabilty criterion, a continuous family of 4 qubit mixed state is constructed as a generalization of Smolin's mixed state in [22], each mixed state in this family is separable under any $2:2$ cut and entangled under any $1:3$ cut, thus they are bound entanglement if 4 parties are isolated (Example 3). Since our invariants can be computed easily and can be used to distinguish inequivalent mixed states under local unitary transformations, it is proved that the {\em generic} members of this continuous family of mixed states are inequivalent under local unitary transformations, thus these 4 qubit mixed states are {\em continuous many} distinct bound entangled mixed states. An example (see Example 2) of {\em continuous many} LOCC-incomparable enatngled tripartite pure states are also constructed to show it is hopeless to characterize the entanglement properties of multipartite pure states by the eigenvalue vectors of their partial traces.\\
  
The algebraic sets used in this paper are  called {\em determinantal varieties} in algebraic geometry ([23] Lecture 9 and [24] Cha.II) and have been studied by mathematicians from different motivations ([23,24--26]). For the algebraic geomerty used in this paper we refer to the nice book [23].\\

We consider the following situation. Alice and Bob share a bipartite quantum system $H_A^m \otimes H_B^n$, and they have a mixed state $\rho$. Now they want to understand the entanglement properties of $\rho$. It is certain that they can prepare any separable pure state  $\phi_1 \otimes \phi_2$ seaprately. Now they measure $\rho$ with this separable pure state, the expectation value is $<\phi_1 \otimes \phi_2|\rho|\phi_1 \otimes \phi_2>$. If Alice's pure state $\phi_1$ is fixed, then $<\phi_1 \otimes \phi_2|\rho|\phi_1 \otimes \phi_2>$ is a Hermitian bilinear form on Bob's pure states (ie., on $H_B^n$). We denote this bilinear form by $<\phi_1|\rho|\phi_1>$. Intuitively the {\em degenerating locus} $V_A^k(\rho)=\{\phi_1 \in P(H_A^m): rank (<\phi_1|\rho|\phi_1>) \leq k\}$ should contain the physical information of $\rho$ and it is almost obvious that these {\em degenerating locus} are invariant under local unitary transformations. Now we use the coordinate form of this formalism. Let $H=H_{A}^m \otimes H_{B}^n$, the standard orthogonal base be $\{|ij>\}$, where, $i=1,...,m$ and $j=1,...,n$, and $\rho$ be a mixed state on $H$. We represent the matrix of $\rho$ in the base $\{|11>,...|1n>,...,|m1>,...,|mn>\}$, and consider $\rho$ as a blocked matrix $\rho=(\rho_{ij})_{1 \leq i \leq m, 1 \leq j \leq m}$ with each block $\rho_{ij}$ a $n \times n$ matrix corresponding to the $|i1>,...,|in>$ rows and the $|j1>,...,|jn>$ columns. For any pure state $\phi_1=r_1|1>+...+r_m|m> \in P(H_A^m)$ the matrix of the Hermitian linear form $<\phi_1|\rho|\phi_1>$ with the base $|1>,...,|n>$ is $\Sigma_{i,j} r_ir_j^{*} \rho_{ij}$. Thus the ``degenerating locus'' are defined as follows.\\

{\bf Definition 1.}{\em  We define 

$$
\begin{array}{ccccc}
V_{A}^k(\rho)=\{(r_1,...,r_m)\in CP^{m-1}:rank( \Sigma_{i,j}r_ir_j^{*} \rho_{ij}) \leq k\}
\end{array}
(1)
$$
for $k=0,1,...,n-1$. Similarly $V_{B}^k (\rho) \subseteq CP^{n-1}$ can be defined. Here * means the conjugate of complex numbers.}\\

{\bf Theorem 1.}{\em  Let $T=U_{A} \otimes U_{B}$, where $U_{A}$ and $U_{B}$ are unitary transformations on $H_{A}^m$ and $H_{B}^n$ rescpectively. Then $V_{A}^k (T(\rho))=U_{A}^{-1}(V_{A}^k(\rho))$.}\\

{\bf Proof.} This is clear from the fact $V_A^k(T(\rho))=\{\phi_1 \in P(H_A^m):rank(U_B<U_A(\phi_1)|\rho|U_A(\phi_1)>(U_B^*)^{\tau}) \leq k \}$.\\

Since $U_{A}^{-1}$ certainly preserve the Fubini-Study metric of $CP^{m-1}$, we know that all metric properties of $V_{A}(\rho)$ are preserved when the local unitary transformations are applied to the mixed state $\rho$.\\

In the following statement, the term {\em algebraic set} means the zero locus of several multi-variable homogeneous polynomials.(see [23]).\\

{\bf Theorem 2.} {\em $V_{A}^k(\rho)$ (resp. $V_{B}^k(\rho)$) is an algebraic set in $CP^{m-1}$ (resp. $CP^{n-1}$).}\\

In the case $\rho=|v><v|$, a pure state in a bipartite quantum system $H_A^m \otimes H_B^n$, it is easy to compute that $k=codim V_A^0(\rho) =m-1-dim V_A^0(\rho)=n-1-dim V_B^0(\rho)=codim V_B^0(\rho)$ is the Schmidt number of the pure state $|v>$.\\

{\bf Theorem 3.} {\em If $\rho$ is a separable mixed state, $V_{A}^k(\rho)$ (resp. $V_{B}^k(\rho)$) is a linear subset of $CP^{m-1}$ (resp. $CP^{n-1}$),ie., it is the sum of the linear subspaces.}\\

For the purpose to prove Theorem 2 and 3 we need some preparation. Let $\{|11>,...,|1n>,...,|m1>,...,|mn>\}$ be the standard orthogonal base of $H=H_A^m \otimes H_B^n$ and $\rho= \Sigma_{l=1}^{t} p_l |v_l><v_l|$ be a mixed state on $H$ with $p_1,...,p_t >0$. Suppose $v_l=\Sigma_{i,j=1}^{m,n} a_{ijl} |ij>$ , $A=(a_{ijl})_{1\leq i \leq m, 1 \leq j \leq n, 1 \leq l \leq t}$ is the $mn \times t$ matrix. Then it is clear that the matrix representation of $\rho$ with the base $\{|11>,...,|1n>,...,|m1>,...,|mn>\}$ is $AP(A^{*})^{\tau}$, where $P$ is the diagonal matrix with diagonal entries $p_1,...,p_t$. As indicated [15], the image of $\rho$ is the linear span of vectors $v_1,...,v_t$. We may consider the $mn\times t$ matrix $A$ as a $m\times 1$ blocked matrix with each block $A_w$, where $w=1,...,m$, a $n\times t$ matrix corresponding to $\{|w1>,...,|wn>\}$. Then it is easy to see $\rho_{ij}=A_iP(A_j^{*})^{\tau}$, where $i=1,...m,j=1,...,m$. Thus\\

$$
\begin{array}{cccccc}
\Sigma r_ir_j^{*} \rho_{ij}=(\Sigma r_i A_i)P(\Sigma r_i^{*} A_i^{*})^{\tau}
\end{array}
(2)
$$

{\bf Lemma 1.} {\em $\Sigma r_ir_j^{*} \rho_{ij}$ is a (semi) positive definite $n\times n$ matrix. Its rank equals to the rank of $(\Sigma r_i A_i)$.}\\

{\bf Proof.} The first conclusion is clear. The matrix $\Sigma r_ir_j^{*} \rho_{ij}$ is of rank $k$ if and only if there exist $n-k$ linear independent vectors $c^j=(c_1^j,...,c_n^j)$ with the property.\\

$$
\begin{array}{cccccccccc}
c^j(\Sigma_{ij}  r_ir_j^{*} \rho_{ij})(c^{j*})^{\tau}=\\
(\Sigma_i r_i c^jA_i)P(\Sigma_i r_i^{*} c^{j*}A_i^{*})^{\tau}=0
\end{array}
(3)
$$

Since $P$ is a strictly positive definite matrix,our conclusion follows immediately.\\

{\bf Proof of Theorem 2.} From Lemma 1 , we know that $V_{A}^k(\rho)$ is the zero locus of all $(k+1)\times (k+1)$ submatrices of $(\Sigma r_iA_i)$. The conclusion is proved.\\

Because the determinants of all $(k+1) \times (k+1)$ submatrices of $(\Sigma r_i A_i)$ are homogeneous polynomials of degree $k+1$ , thus $V_{A}^k (\rho)$(resp. $V_{B}^k(\rho)$) is an algebraic subset (called determinantal varieties in algebraic geometry [23],[24]) in $CP^{m-1}$(resp. $CP^{n-1}$).\\

The point here is: for different representations of $\rho$ as $\rho=\Sigma_j p_j |v_j><v_j|$ with $p_j$'s  positive real numbers, the determinantal varieties from their corresponding $\Sigma_i r_iA_i$'s are the same.\\

Now suppose that the mixed state $\rho$ is separable,ie, there are unit product vectors $a_1 \otimes b_1,....,a_s\otimes b_s$ such that $\rho=\Sigma_{l=1}^{s}q_l |a_l \otimes b_l><a_l \otimes b_l|$ , where $q_1,...q_s$ are positive real numbers. Let $a_u=a_u^1 |1>+...+a_u^m |m>,b_u= b_u^1 |1>+...+b_u^n|n>$ for $u=1,...,s$. Hence the vector representation of $a_u \otimes b_u$ with the standard base is $a_u \otimes b_u= \Sigma_{ij} a_u^ib_u^j |ij>$. Consider the corresponding $mn \times s$ matrix $C$ of $a_1 \otimes b_1,...,a_s \otimes b_s$ as in Lemma 1, we have $\rho=CQ(C^{*})^{\tau}$, where $Q$ is diagonal matrix with diagonal entries $q_1,...,q_s$. As before we consider $C$ as $m\times 1$ blocked matrix with blocks $C_w$, $w=1,...m$. Here $C_w$ is a $n \times s$ matrix of the form $C_w=(a_j^{w}b_j^i)_{1\leq i \leq n, 1 \leq j \leq s}=BT_w$ , where $B=(b_j^i)_{1\leq i \leq n, 1\leq j\leq s}$ is a $n \times s$ matrix and $T_w$ is a diagonal matrix with diagonal entries $a_1^{w},...,a_{s}^{w}$. Thus from Lemma 1, we have $\rho_{ij}=C_i Q (C_j^{*})^{\tau}=B(T_i Q (T_j^{*})^{\tau})(B^{*})^{\tau}=BT_{ij}(B^{*})^{\tau}$, where $T_{ij}$ is a diagonal matrix with diagonal entries $q_1 a_1^i(a_1^j)^{*},...,q_s a_s^i(a_s^{j})^{*}$.\\

{\bf Proof of Theorem 3.} As in the proof of Theorem 2, we have\\

$$
\begin{array}{ccccccc}
\Sigma r_ir_j^{*} \rho_{ij}=\Sigma r_i r_j^{*}BT_{ij}(B^{*})^{\tau}\\
B(\Sigma r_ir_j^{*}T_{ij})(B^{*})^{\tau}
\end{array}
(4)
$$

Here we note  $\Sigma r_ir_j^{*}T_{ij}$ is a diagonal matrix with diagonal entries\\ 
$ q_1(\Sigma r_ia_1^{i})(\Sigma r_ia_1^{i})^{*},...,q_s(\Sigma r_ia_s^{i})(\Sigma r_ia_s^{i})^{*}$.Thus $\Sigma r_ir_j^{*} \rho_{ij}=BGQ(G^{*})^{\tau}(B^{*})^{\tau}$,\\ where $G$ is a diagonal matrix with diagonal entries $ \Sigma r_ia_1^{i},...,\Sigma r_ia_s^{i}$. Because $Q$ is a strictly positive definite matrix, from Lemma 1 we know that  $\Sigma r_ir_j^{*} \rho_{ij}$ is of rank smaller than $k+1$ if and only if the rank of $BG$ is strictly smaller than $k+1$. Note that $BG$ is just the multiplication of $s$ diagonal entries of $G$ (which is linear forms of $r_1,...,r_m$) on the $s$ columns of $B$,  thus the determinants of all $(k+1) \times (k+1)$ submatrices of $BG$ (in the case $ s \geq k+1$, otherwise automatically linear)are the multiplications of a constant (possibly zero) and $k+1$ linear forms of $r_1,...,r_m$. Thus the conlusion is proved.\\

From Lemma 1 and the proof of Theorem 2 and 3 , $\Sigma_i r_iA_i$ play a key role. If we take the standard $\rho= \Sigma_{j=1}^r p_j |\varphi_j><\varphi_j|$, where $p_j , |\varphi_j>, j=1,...,r$ are eigenvalues and eigenvectors, the corresponding $\Sigma_i r_i A_i$ measures the ``geometric position'' of eigenvectors in $H_A^m \otimes H_B^n$. It is obvious from the proof of Theorem 2, the non-local invariants defined in Definition 1 are independent of $p_1,...,p_r$ , the global eigenvalue spectra of the mixed states.\\

Now we introduce the algebraic-geometric invariants of the mixed states and prove the results for tripartite case.  The arbitrary multipartite case is similar and we just generalize directly.\\

Let $H=H_{A}^m \otimes H_{B}^n \otimes H_{C}^l$ and the standard orthogonal base is $|ijk>$, where, $i=1,...,m$,$j=1,...,n$ and $k=1,...,l$, and $\rho$ is a mixed state on $H$. We represent the matrix of $\rho$ in the base $\{|111>,...|11l>,...,|mn1>,...,|mnl>\}$ as
 $\rho=(\rho_{ij,i'j'})_{1 \leq i ,i' \leq m, 1 \leq j ,j' \leq n}$, and  
$\rho_{ij,i'j'}$ is a $l \times l$ matrix. Consider $H$ as a bipartite system as $H=(H_{A}^m \otimes H_{B}^n) \otimes H_{C}^l$, then we have $V_{AB}^k(\rho)=\{(r_{11},...,r_{mn}) \in C^{mn}:rank( \Sigma r_{ij}r_{i'j'}^{*} \rho_{ij,i'j'})\leq k \}$ defined previously. This set is actually the {\em degenerating locus} of the Hermitian bilinear form $<\phi_{12}|\rho|\phi_{12}>$ on $H_C^l$ for the given pure state $\phi_{12} =\Sigma_{i,j}^{m,n} r_{ij} |ij> \in P(H_A^m \otimes H_B^n)$. When the finer cut A:B:C is considered, it is natural to take $\phi_{12}$ as a separable pure state $\phi_{12}=\phi_1 \otimes \phi_2$, ie., there exist $\phi_1=\Sigma_i r_i^1 |i> \in P(H_A^m),\phi_2=\Sigma_j r_j^2 |j> \in P(H_B^n)$ such that $r_{ij}=r_i^1r_j^2$. In this way the tripartite mixed state $\rho$ is measured by tripartite separable pure states $\phi_1 \otimes \phi_2 \otimes  \phi_3$. Thus it is natural we define $V_{A:B}^k(\rho)$ as follows. It is the {\em degenerating locus} of the bilinear form $<\phi_1 \otimes \phi_2 |\rho|\phi_1 \otimes \phi_2>$ on $H_C^l$.\\

{\bf Definition 2.}{\em  Let $\phi:CP^{m-1} \times CP^{n-1} \rightarrow CP^{mn-1}$  be the mapping defined by\\

$$
\begin{array}{ccccccccc}
\phi(r_1^1,...r_m^1,r_1^2,...,r_n^2)=(r_1^1r_1^2,...,r_i^1r_j^2,...r_m^1r_n^2)
\end{array}
(5)
$$

(ie., $r_{ij}=r_i^1 r_j^2$ is introduced.)

Then $V_{A:B}^k (\rho)$ is defined as the preimage $\phi^{-1}(V_{AB}^k(\rho))$.}\\

Similarly $V_{B:C}^k(\rho),V_{A:C}^k(\rho)$ can be defined. In the following statement we just state the result for $V_{A:B}^k(\rho)$. The conclusion holds similarly for other $V's$.\\

From this definition and Theorem 2 we immediately have the following result.\\

{\bf Theorem 2'.} {\em $V_{A:B}^k(\rho)$ is an algebraic set in $CP^{m-1} \times CP^{n-1}$.}\\

{\bf Theorem 1'.}{\em  Let $T=U_{A} \otimes U_{B} \otimes U_{C}$, where $U_{A},U_{B}$ and $U_{C}$ are unitary transformations of $H_A^m, H_B^n, H_C^l$,be a local unitary transformations of $H$. Then $V_{A:B}^k(T(\rho))=U_{A}^{-1} \times U_{B}^{-1}(V_{A:B}^k(\rho))$.}\\

{\bf Proof.} Let $U_{A}=(u_{ij}^{A})_{1 \leq i \leq m,1 \leq j \leq m}$, $U_{B}=(u_{ij}^{B})_{1 \leq i \leq n, 1 \leq j \leq n}$ and $U_{C}=(u_{ij}^{C})_{1 \leq i \leq l, 1 \leq j \leq l}$, be the matrix in the standard orthogonal bases.\\

Recall the proof of Theorem 1, we have $V_{AB}^k(T(\rho))=(U_{A} \times U_{B})^{-1}(V_{AB}^k(\rho))$ under the coordinate change\\

$$
\begin{array}{cccccccc} 
r_{kw}'=\Sigma_{ij} r_{ij} u_{ik}^{A} u_{jw}^{B}\\
=\Sigma_{ij}r_i^1 r_j^2 u_{ik}^{A} u_{jw}^{B}\\
=\Sigma_{ij} (r_{i}^1 u_{ik}^A) (r_{j}^2 u_{jw}^B)\\
=(\Sigma_i r_i^1 u_{ik}^A)(\Sigma_j r_j^2 u_{jw}^B)
\end{array}
(6)
$$

for $k=1,...,m,w=1,...,n$. Thus our conclusion follows from the definition 2.\\

 Since $U_{A}^{-1} \times U_{B}^{-1}$ certainly preserves the  (product) Fubini-Study metric of $CP^{m-1} \times CP^{n-1}$, we know that all metric properties of $V_{A:B}^k(\rho)$ are preserved when the local unitary transformations are applied to the mixed state $\rho$.\\

In the following statement we give a separability criterion of the mixed state $\rho$ under the cut A:B:C. The term {\em a linear subspace of $CP^{m-1} \times CP^{n-1}$} means the product of a linear subspace in $CP^{m-1}$ and a linear subspace in $CP^{n-1}$.\\

{\bf Theorem 3'.}{\em If $\rho$ is a separable mixed state on $H=H_{A}^m \otimes H_{B}^{n} \otimes H_{C}^l$ under the cut A:B:C, $V_{A:B}^k(\rho)$ is a linear subset of $CP^{m-1} \times CP^{n-1}$, ie., it is the sum of the linear subspaces.}\\

{\bf Proof.} We first consider the separability of $\rho$ under the cut AB:C,ie., $\rho= \Sigma_{f=1}^g p_f P_{a_f \otimes b_f}$, where $a_f \in H_{A}^m \otimes H_{B}^n$ and $b_f \in H_{C}^l$ for $f=1,...,g$. Consider the separability of $\rho$ under the cut A:B:C, we have  $a_f=a_f' \otimes a_f''$ , $a_f' \in H_{A}^m, a_f'' \in H_{B}^n$. Let $a_f=(a_f^1,...,a_f^{mn}), a_f'=(a_f'^1,...,a_f'^m)$ and $a_f''(a_f''^1,...,a_f''^n)$ be the coordinate forms with the standard orthogonal basis $\{|ij>\}$, $\{|i>\}$ and $\{|j>\}$ respectively, we have that $a_f^{ij}=a_f'^i a_f''^j$. Recall the proof of Theorem 3, the diagonal entries of $G$ in the proof of Theorem 3 are\\ 

$$
\begin{array}{cccccccc}
\Sigma_{ij}r_{ij} a_f^{ij}=\\
\Sigma_{ij} r_i^1 a_f'^i r_j^2 a_f''^j=\\
(\Sigma_i r_i^1 a_f'^i)(\Sigma_j r_j^1 a_f''^j)
\end{array}
(7)
$$

Thus as argued in the proof of Theorem 3, $V_{A:B}^k(\rho)$ has to be the zero locus of the multiplications of the linear forms in (3). The conclusion is proved.\\

For the mixed state $\rho$ in the multipartite system $H=H_{A_1}^{m_1} \otimes \cdots \otimes H_{A_k}^{m_k}$, we want to study the entanglement under the cut $ A_{i_1}:A_{i_2}:...:A_{i_l}:(A_{j_1}...A_{j_{k-l}})$, where $\{i_1,...,i_l\} \cup \{j_1,...j_{k-l}\}=\{1,...k\}$. We can define the set $V_{A_{i_1}:...:A_{i_l}}^k(\rho)$ similarly. We have the following results.\\

{\bf Theorem 2''.} {\em $V_{A_{i_1}:...:A_{i_l}}^k(\rho)$ is an algebraic set in in $CP^{m_{i_1}-1} \times \cdots \times CP^{m_{i_l}-1}$.}\\

{\bf Theorem 1''.}{\em  Let $T=U_{A_{i_1}} \otimes \cdots \otimes U_{A_{i_l}} \otimes U_{j_1...j_{k-l}}$, where $U_{A_{i_1}},...,U_{A_{i_l}}, U_{j_1...j_{k-l}}$ are unitary transformations of $H_{A_{i_1}},...,H_{A_{i_l}}$, be a local unitary transformations of $H$. Then $V_{A_{i_1}:...:A_{i_l}}^k(T(\rho))=U_{A_{i_1}}^{-1} \times \cdots \times U_{A_{i_l}}^{-1}(V_{A_{i_1}:...:A_{i_l}}^k(\rho))$.}\\

{\bf Theorem 3''.}{\em If $\rho$ is a separable mixed state on $H=H_{A_1}^{m_1} \otimes \cdots \otimes H_{A_k}^{m_k}$ under the cut $ A_{i_1}:A_{i_2}:...:A_{i_l}:(A_{j_1}...A_{j_{k-l}})$, $V_{A_{i_1}:...:A_{i_l}}^k(\rho)$ is a linear subset of $CP^{m_{i_1}-1}  \times ... \times CP^{m_{i_l}-1}$,ie., it is the sum of the linear subspaces.}\\

We now study some examples of mixed states based on our above results.\\

{\bf Example 1.} Consider rank $m$ mixed states $\rho$ in bipartite quantum system $H_A^m \otimes H_B^m$. Thus they have $m$ eigenvectors and from the above description of their $\Sigma_i r_i A_i$'s, they  are $m \times m$ matrices. Hence their $V_A^{m-1}(\rho)$'s are zero locus of {\em one} homogeneous polynomial (ie., the determinant of $\Sigma_i r_i A_i$) in $CP^{m-1}$. In the (most) cases that this polynomial cannot be factorized into $m$ linear forms, thus it is clear that the zero locus $V_A^{m-1}(\rho)$ of this polynomial cannot be linear and the corresponding mixed states are entangled from our criterion Theorem 3. \\

We can consider the following concrete case. Let $H=H_A^3 \otimes H_B^3$ and  $\rho_{t,v,s}=\frac{1}{3}(|v_1><v_1>+|v_2><v_2|+|v_3><v_3|)$ ($t,v,s$ are complex parameters), a mixed state on $H$, where \\

$$
\begin{array}{cccccccc}
v_1=\frac{1}{\sqrt{|t|^6+2}}(t^3 |11>+|22>+|33>)\\
v_2=\frac{1}{\sqrt{|v|^6+2}}(v^3 |12>+|23>+|31>)\\
v_3=\frac{1}{\sqrt{|s|^6+2}}(s^3 |13>+|21>+|32>)
\end{array}
(8)
$$

It is easy to calculate that $\Sigma r_i A_i$ (upto a constant) is the following $3 \times 3$ matrix\\

$$
\left(
\begin{array}{ccc}
t^3r_1&r_3&r_2\\
r_2&v^3r_1&r_3\\
r_3&r_2&s^3r_1
\end{array}
\right)
(9)
$$

Thus $V_{A}^2(\rho_t)$ is defined by $(tvs)^3r_1^3+r_2^3+r_3^3-(t^3+v^3+s^3)r_1r_2r_3=0$ in $CP^{2}$. With $tvsr_1=r_1'$ we have $r_1'^{3}+r_2^3+r_3^3-(\frac{t^3+v^3+s^3}{tvs})r_1'r_2r_3=0$. This universal family of elliptic curves ([27], section 7.2 pp.363---396)is just $V_A^2(\rho_{t,v,s})$.\\

It is easy to check that all global and local spectra (ie., eigenvalues of\\ $\rho_{t,v,s},tr_A(\rho_{t,v,s}),tr_B(\rho_{t,v,s})$) are the same for different parameters with  $|t|^6=|v|^6=|s|^6$ fixed. In this case, set $t^3=he^{i\theta_1},v=he^{i\theta_2},s=he^{i\theta_3}$, where $h>0$, a fixed constant,  we have a family of isospectral (both global and local) mixed states $\rho_{\theta_1,\theta_2,\theta_3}$. The corresponding family of ellptic curves $V_A(\rho_{\theta_1,\theta_2,\theta_3})$ is defined by $(r_1')^3+r_2^3+r_3^3-(\frac{e^{i\theta_1}+e^{i\theta_2}+e^{i\theta_3}}{e^{i(\theta_1+\theta_2+\theta_3)/3}})r_1'r_2r_3=0$. Set $g(\theta_1,\theta_2,\theta_3)=\frac{e^{i\theta_1}+e^{i\theta_2}+e^{i\theta_3}}{e^{i(\theta_1+\theta_2+\theta_3)/3}}$. Let $k(x)=\frac{x^3(x^3+216)^3}{(-x^3+27)^3}$ be the moduli function of elliptic curves. From algebraic-gemetry  we know that the elliptic curve is not the union of 3 lines if $(g(\theta_1,\theta_2,\theta_3))^3 \neq 0,-216,27$ and two elliptic curves corresponding to diffferent paprameters are isomorphic if and only if their corresponding moduli function values are the same (see [27],section 7.2, pp.363-396).\\

{\bf Theorem 4.} {\em $\{\rho_{\theta_1,\theta_2,\theta_3}\}_{\theta_1,\theta_2,\theta_3}$ is a family of entangled mixed state for parameters satisfying  $(g(\theta_1,\theta_2,\theta_3))^3 \neq 0,-216,27$. Moreover $\rho_{\theta_1,\theta_2,\theta_3}$ and $\rho_{\theta'_1,\theta'_2,\theta'_3}$ are not equivalent under local unitary transformations if $k(g(\theta_1,\theta_2,\theta_3)) \neq k(g(\theta'_1,\theta'_2,\theta'_3))$.}\\
 
{\bf Proof.} The conclusion follows from Theorem 3,1 and the above-mentioned fact in algebraic geometry. \\

This can be compared with the example in [20]. In [20] two mixed states $\rho$ and $\delta$ with the same global and local spectra are given, however $\rho$ is entangled and $\delta$ is separable. Thus $\rho$ and $\delta$ are not equivalent under local unitary transformations. Our this example provides  {\em continuous many } inequivalent (under local unitary transformations) entangled mixed states with the same global and local spectra. The {\em continuous many} mixed states in Theorem 4  offer stronger evidence for the point in [20] that a complete understanding of bipartite quantum systems cannnot be obtained by only studying the global and local properties of their spectra.\\

If $|t|^6=|v|^6=|s|^6=1$, the 3 eigenvalues of $\rho,tr_A(\rho),tr_B(\rho)$ are all the same value $\frac{1}{3}$. In these case, the entropy criterion of separability $S(tr_{A}(\rho)), S(tr_B(\rho))$\\$ \leq S(\rho)$ ([10]) and the disorder criterion of separability $\lambda(\rho) \prec \lambda(tr_A(\rho)),\lambda(tr_B(\rho))$ in [20] are all fufilled. However it is easy to see from Theorem 4 that for generic parameters $t,v,s$ with $|t|^6=|v|^6=|s|^6=1$,  the mixed state $\rho_{t,v,s}$ is entangled. Thus our separability criterion can be used to detect entangled mixed states not violating entropy and disorder criteria.\\

In [28] Nielsen gave a beautiful necessary and sufficient condition for the pure state $|\psi>$ can be transformed to the pure state $|\phi>$ in bipartite quantum systems by local operations and classical communications (LOCC) based on the majorization between the eigenvalue vectors of the partial traces of $|\psi>$ and $|\phi>$. In [29] an example was given, from which we know that Nielsen's criterion cannot be generalized to multipartite case, {\bf 3EPR} and {\bf 2GHZ} are understood as pure states in a $4 \times 4 \times 4$ quantum system, they have the same eigenvalue vectors when traced over any subsystem. However it is proved that they are LOCC-incomparable in [29]\\

In the following example a continuous family $\{\phi\}_{\theta_1,\theta_2,\theta_3}$ of pure states in tripartite quantum system $H_{A_1}^3 \otimes H_{A_2}^3 \otimes H_{A_3}^3$ is given, the eigenvalue vectors of $tr_{A_i}(|\phi_{\theta_1,\theta_2,\theta_3}><|\phi_{\theta_1,\theta_2,\theta_3}>), tr_{A_iA_j}(|\phi_{\theta_1,\theta_2,\theta_3}><\phi_{\theta_1,\theta_2,\theta_3}|)$ are independent of parameters $\theta_1,\theta_2,\theta_3$. However the {\em generic} pure states in this family are entangled and  LOCC-incomparable. This gives stronger evidence that it is  hopeless to characterize the entanglement properties of pure states in multipartite quantum systems by only using the eigenvalue spetra of their partial traces.\\

Let $|\phi_{\theta_1,\theta_2,\theta_3}>=\frac{1}{\sqrt{3}}(|v_1> \otimes |1>+|v_2>\otimes |2>+|v_3> \otimes |3>)$. This is a continuous family of pure states in $H$ parameterized by three real parameters. It is clear that $tr_{A_3}|\phi_{\theta_1,\theta_2,\theta_3}><\phi_{\theta_1,\theta_2,\theta_3}|=\frac{1}{3}(|v_1><v_1|+|v_2><v_2|+|v_3><v_3|)$ is a rank 3 mixed state in $H_{A_1}^3 \otimes H_{A_2}^3$. $|\phi_{\theta_1,\theta_2,\theta_3}>$ and $|\phi_{\theta'_1,\theta'_2,\theta'_3}>$ are not equivalent under local unitary transformations if $k(g(\theta_1,\theta_2,\theta_3)) \neq k(g(\theta'_1,\theta'_2,\theta'_3))$,  since their corresponding traces over $A_3$ are not equivalent under local unitary transformations of $H_{A_1}^3 \otimes H_{A_2}^3$ from Theorem 4. Hence the {\em generic} members of this family of pure states in tripartite quum system $H$ are enatngled and  LOCC-incomparable from Theorem 1 in [29]. \\

In the following example we construct a family of rank 7 mixed states $\{\rho_{e_1,e_2,e_3}\}$ ($e_1,e_2,e_3$ are real parameters)  with $\rho_{e_1,e_2,e_3}=\rho_{e_1,e_2,e_3}^{PT}$ (hence PPT automatically) in $H=H_{A}^4 \otimes H_{B}^6$. We prove they are entangled by Theorem 3 (thus bound entanglement) for generic parameters $e_1,e_2,e_3$. This family and the method used here  can be easily generalized to construct entangled mixed states with $\rho=\rho^{PT}$ systematically.\\

{\bf Example 2.} Consider the following 4 $ 6 \times 7$ matrices\\

$$
A_1=
\left(
\begin{array}{ccccccccccc}
1&0&0&0&0&0&0\\
0&1&0&0&0&0&0\\
0&0&1&0&0&0&0\\
0&0&0&2&0&0&1\\
0&0&0&0&2&0&0\\
0&0&0&0&0&2&0
\end{array}
\right)
$$

$$
A_2=
\left(
\begin{array}{ccccccccccc}
0&1&1&-1&0&0&1\\
1&0&1&0&0&0&0\\
1&1&0&0&0&0&0\\
-1&0&0&0&1&1&0\\
0&0&0&1&0&1&0\\
0&0&0&1&1&0&0
\end{array}
\right)
$$

$$
A_3=
\left(
\begin{array}{ccccccccccc}
e_2+e_3&e_1&0&0&0&0&0\\
e_1&e_2&e_3&0&0&0&0\\
0&e_3&e_1+e_2&0&0&0&0\\
0&0&0&e_2+e_3&e_1&0&0\\
0&0&0&e_1&e_2&e_3&0\\
0&0&0&0&e_3&e_1+e_2&0
\end{array}
\right)
$$
,where $e_1,e_2$ are real numbers, and $A_4=(I_6,0)$, where $I_6$ is $6\times 6$ unit matrix.\\

Let $A$ be a $24 \times 7$ matrix with 4 blocks $A_1,A_2,A_3,A_4$ where the 24 rows correspond to the standard base $\{|11>,...,|16>,...,|41>,...,|46>\}$. Let $\rho_{e_1,e_2,e_3}$ be $(1/D)(A(A^{*})^{\tau})$ (where $D$ is a normalizing constant), a mixed state on $H$. It is easy to check that $A_i(A_j^{*})^{\tau}=A_j(A_i^{*})^{\tau}$, hence $\rho_{e_1,e_2,e_3}$ is invariant under partial transposition.\\

Let $\psi_1,...,\psi_7 \in H_A^4 \otimes H_B^6$  be $7$ vectors corresponding to $7$ columns of the matrix $A$. It is clear that the range of $\rho_{e_1,e_2,e_3}$ is the linear span of $\psi_1,..,\psi_7$. When $e_1=e_2=0,e_3=1$, $\psi_2-\psi_3=(|1>+|4>-|2>)\otimes (|2>-|3>)$. Thus there are some separable pure states in the range of $\rho_{0,0,1}$. We will show that $\rho_{0,0,1}$ and $\rho_{e_1,e_2,e_3}$ for generic parameters $e_1,e_2,e_3$ are entangled by our separability criterion Theorem 3.\\

As in the proof of Theorem 2 it is easy to compute $F=r_1A_1+r_2A_2+r_3A_3+r_4A_4$.\\                         

$$
\left(
\begin{array}{ccccccccccc}
u_1&r_2+e_1r_3&r_2&-r_2&0&0&r_2\\
r_2+e_1r_3&u_1'&r_2+e_3r_3&0&0&0&0\\
r_2&r_2+e_3r_3&u_1''&0&0&0&0\\
-r_2&0&0&u_2&r_2+e_1r_3&r_2&r_1\\
0&0&0&r_2+e_1r_3&u_2'&r_2+e_3r_3&0\\
0&0&0&r_2&r_2+e_3r_3&u_2''&0
\end{array}
\right)
(10)
$$
,where $u_1=r_1+r_4+(e_2+e_3)r_3,u_1'=r_1+r_4+e_2r_3$ ,$u_1''=r_1+r_4+(e_1+e_2)r_3$ and $u_2=2r_1+r_4+(e_2+e_3)r_3,u_2'=2r_1+r_4+e_2r_3$,$u_2''=2r_1+r_4+(e_1+e_2)r_3$.\\

We consider the following matrix $F'$ which is obtained by adding the 7-th  column of $F$ to the 4-th column of $F$ and adding $r_2/r_1$ of the 7-th column to the 1st column.\\

$$
\left(
\begin{array}{cccccccccccc}
v_1&r_2+e_1r_3&r_2&0&0&0&r_2\\
r_2+e_1r_3&u_1'&r_2+e_3r_3&0&0&0&0\\
r_2&r_2+e_3r_3&u_1''&0&0&0&0\\
0&0&0&v_2&r_2+e_1r_3&r_2&r_1\\
0&0&0&r_2+e_1r_3&u_2'&r_2+e_3r_3&0\\
0&0&0&r_2&r_2+e_3r_3&u_2''&0
\end{array}
\right)
(11)
$$
,where $v_1=r_1+r_4+(e_2+e_3)r_3+ \frac{r_2^2}{r_1}$,$v_2=3r_1+r_4+(e_2+e_3)r_3$.\\

It is clear that the determinantal varieties defined by $F$ and $F'$ are the same in the affine chart $C^3$ defined by $r_1\neq 0$. Consider the zero locus $Z_1$ defined by the condition that the determinants of the 2 diagonal $3\times3$ submtrices of the 1st $6 \times 6$ submatrix in (11)  are zero, locus $Z_2$ defined by the condition that the 1st 3 rows in (11) are linear dependent and the locus $Z_3$ defined by the condition that the last 3 rows in (11) are linear dependent, it is clear that $V_A^5(\rho_{e_1,e_2,e_3}) \cap C^3$ is the sum of $Z_1,Z_2,Z_3$. We can use the affine coordinates $r_2'=\frac{r_2}{r_1},r_3'=\frac{r_3}{r_1},r_4'=\frac{r_4}{r_1}$ on the affine chart $C^3$ of $CP^3$ defined by $r_1 \neq 0$. In this affine coordinate system all $r_2,r_3,r_4$ should be replced by $r_2',r_3',r_4'$ and $r_1$ should be replaced by 1 in (11). Now we analysis $V_A^5(\rho_{0,0,1})$. It is clear that the following 2 planes $H_1=\{(r_2',r_3',r_4'): r_2'=r_4'+1\},H_2=\{(r_2',r_3',r_4'):r_2'=r_4'+2\}$ are in $V_A^5(\rho_{0,0,1}) \cap C^3$, since in the case $r_2'=r_4'+1$ the 2nd and the 3rd rows of (11) are linearly dependent and in the case $r_2'=r_4'+2$ the 5th and 6th rows of (11) are linearly dependent. The determinants of two $3 \times 3$ diagonal submatrices of the 1st $6 \times 6$ submatrix of (11) are:\\

$$
\begin{array}{cccccccccccccc}
(r_2'-r_4'-1)((r_2')^3 +(r_2')^2r_4'-(r_2')^2 +(r_4')^2+ r_2'r_3'+r_2'r_4'+r_3'r_4'+r_2'+r_3'+2r_4'+1)\\
(r_2'-r_4'-2)((r_4')^2-2(r_2')^2+r_2'r_3'+r_2'r_4'+r_3'r_4'+3r_2'+2r_3'+5r_4'+6)
\end{array}
(12)
$$

Let $X_1$ and $X_2$ be zero locus of the 2nd factors of the above two determinants. It is obvious $X_1 \cap X_2$ is in $V_A^5(\rho_{0,0,1}) \cap C^3$, we want to show that $X_1 \cap X_2 \setminus H_1 \cup H_2$ is a curve, not a line. Take the ponit $P=(0,2,-1) \in X_1 \cap X_2 \cap H_1$, the tangent plane $H_3$ of $X_2$ at $P$ is defined by $4r_2'+r_3'+5r_4'=-3$. If $X_1 \cap X_2$ is a line around the ponit $P$, this line is contained in $H_3 \cap X_2$. However we can easily find that $H_3 \cap X_2$ is defined by $3(r_2')^2+2(r_4')^2+4r_2'r_4'+4r_4'=0$. This polynomial is irreducible and thus $H_3 \cap X_2$ is a curve around the piont $P$. Thus $X_1 \cap X_2$ is a curve around the point $P$. It is easy to check that $X_1 \cap X_2$ is not contained in $H_1$ around the point $P$.   This implies that $V_A^5(\rho_{e_1,e_2,e_3}) \cap C^3$ (actually the locus $Z_1$) contains a curve (not a line) for generic parameters $e_1,e_2,e_3$(including parameters $0,0,1$) from algebraic geometry. Thus if $V_A^5 (\rho_{e_1,e_2,e_3}) \cap C^3$ is the sum of (affine) linear subspaces, it have to contain a dimension 2 affice linear subspace $H_4$ other than $H_1$ and $H_2$ of the affine chart $C^3$. Thus the determinants of all $6 \times 6$ submatrices of (11) have to contain an (fixed) affine linear form (ie., a degree one polynomial of $r_2',r_3',r_4'$ which may contain a constant term) other than $r_2'-r_4'-1$ and $r_2'-r_4'-1$ as one of their factors. This affine linear form defines that dimension 2 linear affine subspace $H_4$ of $C^3$. However it is easy to check this is impossible for generic parameters $e_1,e_2,e_3$ (including parameters $0,0,1$). We know that $V_A^5(\rho_{e_1,e_2,e_3}) \cap C^3$ cannot be the sum of (affine) linear subspaces of $C^3$ for generic $e_1,e_2,e_3$ (including parameters $0,0,1$).  Thus from Theorem 3 $\rho_{e_1,e_2,e_3}$ is entangled for generic parameters $e_1,e_2,e_3$ (including parameters $0,0,1$).\\

From the construction in Example 2 we can see if $A_1,...,A_m$ are $m$ $n \times t$ matrices satisfying $A_i (A_j^*)^{\tau}=A_j (A_i^*)^{\tau}$, $A$ is the $m \times 1$ matrix with $i-th$ block $A_i$ and the rows of $A$ correspond to the base $|11>,...,|1n>,...,|m1>,...,|mn>$ of $H_A^m \otimes H_B^n$, then the mixed state $\rho=\frac{1}{D}A (A^*)^{\tau}$, where $D$ is a normalized constant, is invariant under partial transpose. It is not very difficult to find such matrices. For the purpose that the constructed mixed state $\rho$ is entangled (thus a bound entangled mixed state),  we just need that the determinantal variety $\{(r_1,...,r_m): rank (\Sigma r_i A_i) \leq n-1\}$ is NOT {\em linear}. We know from algebraic-geometry (see [23],[24]) it is not very  hard to find such matrices $A_1,...,A_m$. However as illustrated in Example 2 we do need some explicit calculation to prove this point. Thus our separability criterion and method in Example 2 offer a new systematic way to construct PPT bound entangled mixed states.\\

The following example, which is a continuous family (depending on 4 parameters) of  mixed state in the four-party quantum system $H_A^2 \otimes H_B^2 \otimes H_C^2 \otimes H_D^2$ and separable for any $2:2$ cut but entangled for any $1:3$ cut,  can be thought as a generalization of Smolin's mixed state in [22].\\

{\bf Example 3.} Let $H=H_{A}^2 \otimes H_{B}^2 \otimes H_{C}^2 \otimes H_{D}^2$ and $h_1,h_2,h_3,h_4$ (understood as row vectors)are 4 mutually orthogonal unit vectors in $C^4$. Consider the $16 \times 4$ matrix $T$ with 16 rows as\\
 $T=(a_1h_1^{\tau},0,0,a_2 h_2^{\tau},0, a_3 h_3^{\tau},a_4 h_4^{\tau},0,0, a_5 h_3^{\tau}, a_6 h_4^{\tau},0, a_7 h_1^{\tau},0,0,a_8 h_2^{\tau})^{\tau}$. Let \\ $\phi'_1,\phi'_2,\phi'_3,\phi'_4$ be 4 vectors in $H$ whose expansions with the base $|0000>,|0001>,|0010>,|0011>,|0100>,|0101>,|0110>,|0111>,|1000>$,\\$|1001>,|1010>,|1011>,|1100>,
|1101>,|1110>,|1111> $ are exactly the 4 columns of the matrix $T$ and $\phi_1,\phi_2,\phi_3,\phi_4$ are the normalized unit vectors of $\phi'_1,\phi'_2, \phi'_3, \phi'_4$. Let $\rho=\frac{1}{4}(P_{\phi_1} +P_{\phi_2} +P_{\phi_3} +P_{\phi_4})$.\\

It is easy to check that when $h_1=(1,1,0,0),h_2=(1,-1,0,0), h_3=(0,0,1,1), h_4=(0,0,1,-1)$ and $a_1=a_2=a_3=a_4=1$. It is just the Smolin's mixed state in [22].\\

Now we prove that $\rho$ is invariant under the partial transposes of the cuts AB:CD,AC:BD,AD:BC.\\

Let  the ``representation'' matrix $T=(b_{ijkl})_{i=0,1,j=0,1,k=0,1,l=0,1}$ is the matrix with columns corresponding the expansions of $\phi_1,\phi_2,\phi_3,\phi_4$.Then we can consider that $T=(T_1,T_2,T_3,T_4)^{\tau}$ is blocked matrix of size $4 \times 1$ with each block $T_{ij}=(b_{kl })_{k=0,1,l=0,1}$ a $4 \times 4$ matrix,where $ij=00,01,10,11$. Because $h_1,h_2,h_3,h_4$ are mutually orthogonal unit vectors we can easily check that $T_{ij} (T_{i'j'}^{*})^{\tau}=T_{i'j'} (T_{ij}^{*})^{\tau}$ Thus it is invariant when the partial transpose of the cut AB:CD is applied.\\

With the same methods we can check that $\rho$ is invariant when the partial transposes of the cuts AC:BD, AD:BC  are applied. Hence $\rho$ is PPT under the cuts AB:CD, AC:BD,AD:BC. Thus from a result in [30] we know $\rho$ is separable under these cuts AB:CD, AC:BD,AD:BC.\\

Now we want to prove $\rho$ is entangled under the cut A:BCD by computing $V_{BCD}^1(\rho)$. From the previous arguments , we can check that $V_{BCD}^1(\rho)$ is the locus of the condition: $a_1 h_1 r_{000} + a_2 h_2 r_{011} +a_3 h_3 r_{101} +a_4 h_4 r_{110}$ and $a_7 h_1 r_{100} + a_8 h_2 r_{111} +a_5 h_3 r_{001} +a_6 h_4 r_{010}$ are linear dependent. This is equivalent to the condition that the matrix (12) is of  rank 1.\\

$$
\left(
\begin{array}{cccccc}
a_7 r_{100} & a_8 r_{111} & a_5 r_{001} & a_6 r_{010}\\
a_1 r_{000} & a_2 r_{011} & a_3 r_{101} & a_4 r_{110}
\end{array}
\right)
(12)
$$

From [23] pp. 25-26 we can check that $V_{BCD}^1(\rho)$ is exactly the famous Segre variety in algebraic geometry. It is irreducible and thus cannot be linear. From Theorem 3, $\rho$ is entangled under the cut A:BCD. Similarly we can prove that $\rho$ is entangled under the cuts B:ACD, C:ABD, D:ABC.\\

Now we compute $V_{A:B}^3(\rho)$. From the previous arguments  and Definition 2 , it is just the locus of the condition that the vectors $h_1(a_1 r_0^1 r_0^2 +a_7 r_1^1 r_1^2)$, $h_3 (a_3 r_0^1 r_1^2 +a_5 r_1^1 r_0^2)$, $h_4(a_4 r_0^1 r_1^2 +a_6  r_1^1 r_0^2)$, $h_2 (a_2 r_0^1 r_0^2 +a_8 r_1^1 r_1^2)$ are linear dependent. Since $h_1,h_2,h_3,h_4$ are mutually orthogonal unit vectors,we have \\

$$
\begin{array}{cccccccccc}
V_{A:B}^3(\rho)=\{(r_0^1,r_1^1,r_0^2,r_1^2) \in CP^1 \times CP^1:\\
(a_1 r_0^1 r_0^2 +a_7 r_1^1 r_1^2)(a_3 r_0^1 r_1^2 +a_5 r_1^1 r_0^2)(a_4 r_0^1 r_1^2 +a_6  r_1^1 r_0^2)(a_2 r_0^1 r_0^2 +a_8 r_1^1 r_1^2)=0\}
\end{array}
$$

From Theorem 3 we know that $\rho$ is entangled for the cut A:B:CD, A:C:BD and A:D:BC for generic parameters, since (for example) $a_1r_0^1r_0^2+a_7r_1^1r_1^2$ cannot be factorized to 2 linear forms for generic $a_1$ and $a_7$.  This provides another proof the mixed state is entangled if the 4 parties are isolated.\\
Let $\lambda_1=-a_1/a_7,\lambda_2=-a_3/a_5, \lambda_3=-a_4/a_6, \lambda_4=-a_2/a_8$ and consider the family of the mixed states $\{\rho_{\lambda_{1,2,3,4}}\}$, we want to prove the following statement.\\

{\bf Theorem 5.} {\em The generic memebers in this continuous family of mixed states are inequivalent under the local operations on $H=H_{A}^2 \otimes H_{B}^2 \otimes H_{C}^2 \otimes H_{D}^2$.}\\

{\bf Proof.} From the above computation, $V_{A:B}^3(\rho_{\lambda_{1,2,3,4}})$ is the union of the following 4 algbraic varieties in $CP^1 \times CP^1$.\\

$$
\begin{array}{ccccccccccc}
V_1=\{(r_0^1,r_1^1,r_0^2,r_1^2) \in CP^1 \times CP^1:r_0^1 r_0^2 - \lambda_1 r_1^1 r_1^2=0\}\\
V_2=\{(r_0^1,r_1^1,r_0^2,r_1^2) \in CP^1 \times CP^1:r_0^1 r_1^2 - \lambda_2 r_1^1 r_0^2=0\}\\
V_3=\{(r_0^1,r_1^1,r_0^2,r_1^2) \in CP^1 \times CP^1:r_0^1 r_1^2 - \lambda_3 r_1^1 r_0^2=0\}\\
V_4=\{(r_0^1,r_1^1,r_0^2,r_1^2) \in CP^1 \times CP^1:r_0^1 r_0^2 - \lambda_4 r_1^1 r_1^2=0\}
\end{array}
$$

 From Theorem 1', if $\rho_{\lambda_{1,2,3,4}}$ and $\rho_{\lambda'_{1,2,3,4}}$ are equivalent by a local operation, there must exist 2 fractional linear transformations $T_1, T_2$ of $CP^1$ such that $T=T_1 \times T_2$ (acting on  $CP^1 \times CP^1$) transforms the 4 varieties $V_1,V_2,V_3,V_4$  of $\rho_{\lambda_{1,2,3,4}}$ to the  4 varieties $V'_1,V'_2,V'_3,V'_4$  of $\rho_{\lambda'_{1,2,3,4}}$,ie., $T(V_i)=V'_j$.\\

Introduce the inhomogeneous coordinates $x_1=r_0^1/r_1^1,x_2=r_0^2/h_1^2$. Let $T_1(x_1)=(ax_1+b)/(cx_1+d)$.  Suppose $T(V_i)=V'_i, i=1,2,3,4$. Then we have $ab \lambda_1= cd \lambda'_1 \lambda'_2$ and $ab \lambda_4=cd \lambda'_3 \lambda'_4$. Hence $\lambda_1 \lambda'_3 \lambda'_4 =\lambda'_1 \lambda'_2 \lambda_4$. This means that there are some algebraic relations of parameters if the $T$ exists. Similarly we can get the same conclusion for the other possibilities $T(V_i)=V'_j$. This implies that there are some algebraic relations of parameters $\lambda_{1,2,3,4}$ and $\lambda'_{1,2,3,4}$ if $\rho_{\lambda_{1,2,3,4}}$ and $\rho_{\lambda'_{1,2,3,4}}$ are  equivalent  by a local operation. Hence our conclusion follows immediately.\\

In conclusion, from the consideration of measuring the mixed states by separable pure states we naturally constructed non-local invariants for both bipartite and multipartite mixed states only depending on their eigenvectors via algebraic-geometry of determinantal varieties. These algebraic-gemetric invariants of mixed states can be used to distinguish {\em inequivalent} classes of states under local unitary transformations {\em effectively} as shown in Example 1 and Example 3. We have given an {\em eigenvalue-free } separability criterion based on these algebraic-geometric invariants of mixed states which can detect entangled mixed states not violating entropy and disorder criteria as shown in Example 1. This criterion also works well to detect entangled mixed states for which there are some separable pure states within the their ranges as shown in Example 2. Based on this criterion a new systematic way to construct PPT bound entangled mixed states have been described. The work here strongly indicates that quantum entanglement has a large {\em eigenvalue-free} part which depends only on the position of eigenvectors  and we have developed a {\em quantitative} method to treat it. \\

The author acknowledges the support from NNSF China, Information Science Division, grant 69972049.\\

e-mail: chenhao1964cn@yahoo.com.cn\\

\begin{center}
REFERENCES
\end{center}

1.A.Einstein, B.Podolsky and N.Rosen, Phys. Rev. 47,777(1935)\\

2.E.Schrodinger, Proc.Camb.Philos.Soc.,31,555(1935)\\

3.R.Jozsa, in The Geometric Universe, edited by S.Huggett, L.Mason, K.P.Tod, S.T.Tsou, and N.M.J.Woodhouse (Oxford Univ. Press, 1997)\\

4.A.Ekert, Phys Rev.Lett.67, 661(1991)\\

5.C.H.Bennett, G.Brassard, C.Crepeau, R.Jozsa, A.Peres and W.K.Wootters, Phys.Rev.Lett 70, 1895 (1993)\\

6.C.H.Bennett, G.Brassard, S.Popescu, B.Schumacher, J.Smolin and W.K.Wootters, Phys. Rev.Lett. 76, 722(1996)\\

7.S.Popescu and D.Rohrlich, Phys.Rev.A 56,3319(1997)\\

8.N.Linden, S.Popescu and A.Sudbery, Phys.Rev.Lett. 83,243(1999)\\

9.A.Peres, Quantum Theory:Concepts and Methods (Kluwer, Dordrecht, 1993)\\

10.M.Horodecki, P.Horodecki and R.Horodecki, in Quantum Information--Basic concepts and experiments, edited by G.Adler and M.Wiener (Springer Berlin, 2000)\\

11.A.Peres, Phys.Rev.Lett.,77,1413(1996)\\

12.M.Horodecki,P.Horodecki and R.Horodecki, Phys. Lett.A 223,8(1996)\\

13.R.F.Werner and M.M.Wolf, Phys.Rev. A,61,062102 (2000)\\

14.M.Horodecki, P.Horodecki and R.Horodecki, Phys.Rev.Lett. 80, 5239(1998)\\

15.P.Horodecki, Phys Lett A 232 (1997)\\

16.C.H.Bennett, D.P.DiVincenzo, T.Mor,P.W.Shor, J.A.Smolin and T.M. Terhal, Phys Rev. Lett. 82, 5385 (1999)\\

17.D.DiVincenzo, T.Mor, P.Shor, J.Smolin and B.M.Terhal, Commun. Math. Phys. (accepted) quant-ph/9908070\\

18.D.Bruss and  A.Peres, Phys. Rev. A 61 30301(R) (2000)\\

19.P.Horodecki and M.Lewenstein, Phys. Rev. Lett. 85, 2657 (2000)\\

20.M.A.Nielsen and J.Kempe, Phys. Rev.Lett. 86, 5184 (2001)\\

21.M.Horodecki, P.Horodecki and R.Horodecki, quant-ph/0006071, to appear Phys. Lett.\\

22.J.A.Smolin,Phys.Rev. A 63, 032306(2001)\\

23.J.Harris, Algebraic Geometry, A First course, GTM 133, Springer-Verlag 1992\\

24.E.Arbarello, M.Cornalba, P.A.Griffiths and J.Harris, Geometry of algebraic curves, Vol. I, Springer-Verlag, 1985\\

25.P.A.Griffiths, Compositio Math., 50, 267(1983)\\

26.C.De Concini, D.Eisenbud and C.Procesi, Invent. Math., 56 129(1980)\\

27.E.Brieskorn and H.Knorrer, Ebene algebraische Kurven,Birkhauser, Basel-Boston-Stuttgart,1981\\

28.M.A.Nielsen, Phys. Rev. Lett. 83,436(1999)\\

29.C.H.Bennett,S.Popescu,D.Rohrlich,J.A.Smolin and A.Thapliyal, Phys. Rev. A 63, 012307, 2000\\

30.M.Lewenstein,D.Bruss,J.I.Cirac,B.Krus,J.Samsonowitz,A.Sanpera and R.Tarrach,J.Mod.Optics,47,2481 (2000),quant-ph/0006064\\

\end{document}